\documentclass[pre,floats,twocolumn,superscriptaddress,usenames,dvipsnames]{revtex4-1}

\usepackage{graphicx}
\usepackage{rotate}
\usepackage{epsfig}
\usepackage{xcolor}
\usepackage[normalem]{ulem}
\usepackage{amssymb,amsfonts,amsmath}
\usepackage{dcolumn}
\usepackage{hyperref}
\usepackage{bm}
\usepackage{layouts}

\newcommand{\avg}[1]{\langle #1 \rangle}

\newcommand{\etal}{\emph{et al.} }
\newcommand{\ie}{\emph{i.e.} }
\newcommand{\eg}{\emph{e.g.} }

\graphicspath{{pictures/}}

\begin{document}
	
\title{Interplay between cost and benefits triggers nontrivial vaccination uptake}

\author{Benjamin Steinegger}
\affiliation{Laboratory for Statistical Biophysics, \'Ecole Polytechnique F\'ed\'erale de Lausanne (EPFL), CH-1015 Lausanne, Switzerland}

\author{Alessio Cardillo}
\affiliation{Institut Catal\`a de Paleoecologia Humana i Evoluci\'o Social (IPHES), E-43007 Tarragona, Spain}
\affiliation{GOTHAM Lab -- Institute for Biocomputation and Physics of Complex Systems (BIFI), University of Zaragoza, E-50018 Zaragoza, Spain}

\author{Paolo De Los Rios}
\affiliation{Laboratory for Statistical Biophysics, \'Ecole Polytechnique F\'ed\'erale de Lausanne (EPFL), CH-1015 Lausanne, Switzerland}

\author{Jes\'us G\'omez-Garde\~nes}
\affiliation{Department of Condensed Matter Physics, University of Zaragoza, E-50009 Zaragoza, Spain}
\affiliation{GOTHAM Lab -- Institute for Biocomputation and Physics of Complex Systems (BIFI), University of Zaragoza, E-50018 Zaragoza, Spain}

\author{Alex Arenas}
\affiliation{Department d'Enginyeria Inform\'atica i Matem\'atiques, Universitat Rovira i Virgili, E-43007 Tarragona, Spain}

\date{\today}

\begin{abstract}
The containment of epidemic spreading is a major challenge in science. Vaccination, whenever available, is the best way to prevent the spreading, because it eventually immunizes individuals. However, vaccines are not perfect, and total immunization is not guaranteed. Imperfect immunization has driven the emergence of anti-vaccine movements that totally alter the predictions about the epidemic incidence. Here, we propose a mathematically solvable mean-field vaccination model to mimic the spontaneous adoption of vaccines against influenza-like diseases, and the expected epidemic incidence. The results are in agreement with extensive Monte Carlo simulations of the epidemics and vaccination co-evolutionary processes. Interestingly, the results reveal a non-monotonic behavior on the vaccination coverage, that increases with the imperfection of the vaccine and after decreases. This apparent counterintuitive behavior is analyzed and understood from stability principles of the proposed mathematical model.
\end{abstract}

\maketitle

%
\section{Introduction}
The quantitative study of diseases propagation has captured the attention of statistical physicists since long \cite{anderson-nature-1979,anderson-book-1992,pastor_satorras-rev_mod_phys-2015}. Specifically, this approach has shed light on many conundrums by considering the networked structure of contacts, their time-varying and multi-layer character, as well as the recurrent nature of mobility patterns \cite{pastor_satorras-prl-2001,vazquez-prl-2007,de_domenico-nat_phys-2016,holme-pre-2017,gomez-gardenes-nphys-2018}. Vaccination, whenever possible, is the most effective way to harness and prevent the spreading of a disease \cite{wadman-science-2017}. Under normal circumstances, the decision of getting vaccinated can be considered as an act of \emph{cooperation}, since it bestows benefits on the whole population at the expenses of single individuals. Notwithstanding, we are recently witnessing the emergence of widespread anti-vaccine movements, which are mainly fueled by misconceptions and mischievous news about vaccines \cite{wakefield-lancet-1998,larson-vaccine-2014,nat_microbio-2017,schmid-pone-2017,salmon-arch_ped-2005,brunson-pediatrics-2013,freed-pediatrics-2011}. Scientists (including physicists) are devoting tremendous efforts in designing efficient immunization strategies \cite{wang-physrep-2016} as well as shedding light on the mechanisms behind the deliberate decision of not getting vaccinated and their harmful consequences \cite{betsch-nat_microbio-2017,betsch-nat_humbeh-2017}.\newline
\indent Vaccination can be modeled as a strategy of a game. Under such premises, the evolution of vaccines' voluntary adoption can be investigated using the machinery of game theory \cite{nowak-book-2006,gintis-book-2009} and statistical physics \cite{szabo_hauert,perc}. The first studies carried out on vaccination games use classical game theory, with single round games in which agents have perfect knowledge of their odds to get infected \cite{bauch-pnas-2003,bauch-pnas-2004}. In reality, however, individuals are not perfectly aware of the risk to get infected, and vaccination coverage may evolve in time as byproduct of personal experiences or imitation. Therefore, evolutionary game theory is the natural workbench to tackle the problem. The seminal works in this direction assumed the simultaneous evolution of vaccination and spreading dynamics \cite{bhattacharyya-j_theo_bio-2010,bauch-plos_comp_bio-2012,d_onofrio-pone-2012}. A different approach was used in the case of seasonal influenza, by considering that the spreading process reaches its stationary state before vaccination games take place \cite{fu-proc_roy_soc_b-2011,zhang-physa-2012,cardillo-pre-2013,wu-pone-2011}. \newline
\indent Here, we introduce a mean-field framework to mimic the spontaneous adoption of vaccines against influenza-like diseases. Our model captures the essence of previous approaches to gauge, analytically, the risk of epidemic outbreaks. In particular, we use a minimal evolutionary vaccination game to infer the strategies adopted before an epidemic season. This, in turns, allows us to estimate the risk of future outbreaks by encapsulating agents' strategies into an epidemic model. We get analytical results and insightful conclusions from the analysis of this framework in well-mixed populations. More specifically, we unveil how the interplay between the probability of infection, vaccine effectiveness, and cost, gives rise to non-linear responses in vaccine uptake. Our analysis reveals a non-monotonic behavior on the vaccination coverage which, surprisingly, increases as the vaccine quality deteriorates. Such counterintuitive behavior is analyzed and understood from stability principles of the proposed mathematical model. \newline
%
%

%
\section{The model}
\indent Let us consider a well-mixed population of $N$ agents/individuals. We asume that this population has initially undergone a Susceptible-Infected-Recovered (SIR) disease spreading \cite{anderson-nature-1979,anderson-book-1992}. To stay simple and keep the problem still analytically solvable, we set the disease incidence in such previous season equal to $\alpha$. Such quantity is an input parameter of the model and acts as a proxy for the perception of infection risk, \eg advised through the mass-media. After the first outbreak, the vaccination dynamics takes place. As a result, agents converge to the decision of vaccinating ($V$) or not ($NV$). The resulting strategy is the outcome of the evolutionary dynamics (see below) given the previous incidence, $\alpha$, infection probability, $\beta$, recovery cost, $T$, the cost of the vaccine, $c$, and its failure rate $\gamma$ or, equivalently, its {\em effectiveness} ($1-\gamma$). The corresponding vaccine coverage -- given by the fraction of vaccinated agents $y_{\text{eq}}$ -- is used as the input of a new SIR spreading process, having the same $\beta$ and $T$. The results of the SIR dynamics allow us to estimate the total infection incidence, $R_{\infty}$, as a function of the relevant parameters, especially $\alpha$. The mathematical definition of $\alpha$ is the probability of infection, in the previous outbreak, of an agent without protection against the disease.
\newline
\indent Vaccination dynamics consists of a repeatedly played two strategy game, in which agents either take the vaccine ($s = V$) or not ($s = NV$). Agents will decide according to: {\em (i)} the cost associated to uptaking the vaccine ($c$) and {\em  (ii)} the recovery cost ($T$) weighted by the perceived risk of getting infected by a contact in a future outbreak. The latter risk depends on the previous incidence $\alpha$, the infection rate $\beta$, and the vaccine effectivenesss $(1-\gamma)$. It is worth stressing that the vaccination cost should not be seen purely as a financial one. It also includes for example vaccine hesitancy \cite{larson-vaccine-2014,schmid-pone-2017}. The payoffs associated to each of the four possible encounters between pairs of agents are:
\begin{equation}
\label{eq:Payoff}
\begin{cases}
P_{\text{\tiny{V$\rightarrow$ V}}} &= -c -\gamma^{2} \beta \alpha T \\
P_{\text{\tiny{V$\rightarrow$ NV}}} &= -c -\gamma \beta \alpha T  \\
P_{\text{\tiny{NV$\rightarrow$ V}}} &= -\gamma \beta \alpha  T  \\
P_{\text{\tiny{NV$\rightarrow$ NV}}} &= -\beta \alpha T
\end{cases}
\,.
\end{equation}
Where $P_{s_1 \rightarrow s_2}$ is the payoff accumulated by an agent with strategy $s_1$ when it meets another having strategy $s_2$ ($s_1,s_2 \in \{V,NV\}$). As explained above, the prefactors of the recovery cost ($T$) in Eq.(\ref{eq:Payoff}) are modulated by the perceived risk of infection given the previous outbreak, encapsulated in $\alpha$, and the probability that an individual playing with strategy $s_1$ is infected by another with strategy $s_2$. 

The fitness of an agent $i$, $\pi_{i}$, is defined as the sum of payoffs accumulated across its pairwise interactions. Every agent $i$ with strategy $s_{i}$ chooses randomly another one $j$, with strategy $s_{j}$, compares their fitness ($\pi_{i}, \pi_{j}$) and if $\pi_{j} > \pi_{i}$ adopts $j$'s strategy with probability:
\begin{equation}
\label{eq:updateRule}
\Gamma_{s_{j}\rightarrow s_{i}} = \frac{\pi_{j}-\pi_{i}}{\max\limits_{\forall s_a, s_b, s_c, s_d\in \{V,NV\} }(P_{s_a \rightarrow s_b}-P_{s_c \rightarrow s_d})} \,.
\end{equation}
The strategies of the agents are updated synchronously after each round of the game. Using the above update function, and assuming that agents are well-mixed, the mean outcome of the individual decisions follows the so-called \emph{replicator equation} \cite{nowak-book-2006}:
\begin{equation}
\label{eq:replicator}
\dot{y} = y \, \Bigl(\avg{\pi_{V}} - \avg{\pi} \Bigr) \,,
\end{equation}
where $y$ is the fraction of vaccinated agents, $\avg{\pi_{V}}$ the average payoff of a vaccinated agent, and $\avg{\pi}$ the average payoff of the population. According to the payoffs in Eq.\eqref{eq:Payoff}, the replicator equation reads:
\begin{equation}
\label{eq:diff}
\dot{y} = -y(1-y) \biggl\{ c-\beta \alpha T (1-\gamma) \Bigl[ 1-(1-\gamma)y  \Bigr] \biggr\} \,.
\end{equation}
The above equation admits two trivial equilibrium points ($y_{\text{eq}}=0$ and $y_{\text{eq}}=1$), and a third, nontrivial, one:
\begin{equation}
\label{eq:eqPoint}
y^{*} = \frac{1}{1-\gamma}\left[ 1-\frac{c}{\beta T \alpha (1-\gamma)}\right] \,.
\end{equation}

The criteria for the existence of the non trivial equilibrium point have an intuitive interpretation. In fact, the condition $y^{*} > 0$ is equivalent to $P_{\text{\tiny{NV$\rightarrow$ NV}}} < P_{\text{\tiny{V$\rightarrow$ NV}}}$, which translates into a vaccination threshold $\tilde{\beta} = c/\alpha T (1-\gamma)$. In other words, it is impossible to observe vaccination in a system where $c > T$, \ie where vaccinating costs more than recovery. The criterion $y^{*} < 1$ translates into $P_{\text{\tiny{V$\rightarrow$ V}}} < P_{\text{\tiny{NV$\rightarrow$ V}}}$, corresponding to $\beta < \tilde{\beta}/\gamma$. If this condition is not fulfilled, a vaccinated agent has a higher payoff than a non vaccinated one regardless of the opponents strategies, hence leading to full vaccination. In the next section, we analyze the stability of the equilibrium points.

%
\subsection{Stability analysis of the model}
\label{sec:stability}

The evolution of $y$ is ruled by Eq.~\eqref{eq:diff}. To analyze the stability of its fixed points, we need to evaluate the derivatives of the selection gradient, $F(y)$, corresponding to the RHS of Eq.~\eqref{eq:diff}:
\begin{eqnarray}
\label{eq:derDiff}
\frac{d F(y)}{dy} &= -(1-2y)\biggl\{ c-\beta \alpha T(1-\gamma)\Bigl[1-(1-\gamma)y\Bigr] \biggr\} \nonumber\\
&-y(1-y)\beta \alpha T(1-\gamma)^{2}.
\end{eqnarray}
Evaluating the derivative at the internal fixed point, $y^{*}$, we get:
\begin{equation}
\label{eq:DerInternal}
\frac{d F(y)}{dy} \Bigr|_{y=y^{*}} = -y^{*}(1-y^{*})\beta \alpha T (1-\gamma)^{2}.
\end{equation}
If the internal fixed point lies in ($0 < y^{*}<1$), it is stable, since $\frac{d F(y)}{dy} \Bigr|_{y=y^{*}} <0$. Otherwise, if ($y^{*} < 0$) or ($y^{*} > 1$), the fixed point $ y^{*} $ is unstable. Evaluating the derivative at the monomorphic states ($y_{\text{eq}} = 0$ and $y_{\text{eq}}=1$), we get:
\begin{equation}
\label{eq:DerMon}
\begin{cases}
\frac{d F(y)}{dy} \Bigr|_{y=0} &= \beta \alpha T (1-\gamma) - c = P_{\text{\tiny{V $\rightarrow$ NV}}}-P_{\text{\tiny{NV $\rightarrow$ NV}}} \\
\frac{d F(y)}{dy} \Bigr|_{y=1} &= c-\beta \alpha T(1-\gamma)\gamma = P_{\text{\tiny{NV $\rightarrow$ V}}}-P_{\text{\tiny{V $\rightarrow$ V}}}
\end{cases}\,.
\end{equation}

\begin{figure}[h!]
\centering
\includegraphics[width=1.0\columnwidth]{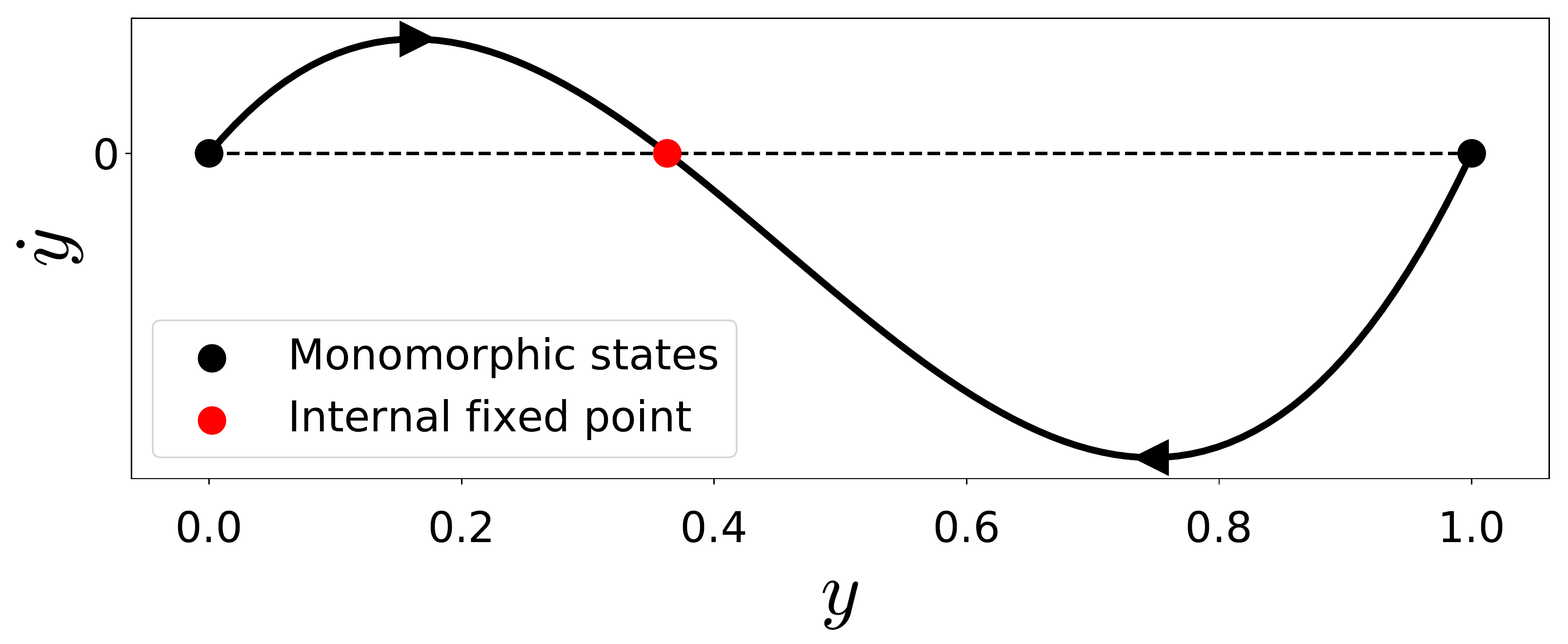}
\caption{Selection gradient, $\dot{y}$ = F(y), as a function of the vaccination coverage, $y$. The parameters are set to $c=0.1$, $\alpha=0.3$, $\beta = 0.55$, $\gamma = 0.1$, and $T=1.0$.}
\label{fig:selectionGradient}
\end{figure}

Consequently, the derivative evaluated at the monomorphic states corresponds to the existence criteria $y^{*} > 0$ and $y^{*} < 1$, respectively. Therefore, the monomorphic states are unstable in the presence of the internal fixed point, $y^{*}$, and reach stability depending on wheter $y^{*} <0$ or $y^{*} > 1$. These findings are summarized in Fig.~\ref{fig:selectionGradient}, which presents the selection gradient as a function of the vaccination coverage, $y$, in the presence of the internal fixed point, $y^{*}$. One can see that $y^{*}$ is stable and as the internal fixed point is shifted to $y^{*} = 0$ or $y^{*} = 1$, the corresponding monomorphic state reaches stability. 

%
%
Therefore, the vaccination coverage, $y_{\text{eq}}$, being defined by the stable equilibrium point, is given by the internal fixed point $y_{\text{eq}} = y^{*}$ for $\tilde{\beta} \leq \beta \leq \tilde{\beta}/\gamma$, whereas for $ \beta < \tilde{\beta}$ and $ \tilde{\beta}/\gamma < \beta$ it is given by $y_{\text{eq}} = 0$ and $y_{\text{eq}} = 1$, respectively.

%
\subsection{Interpretation of results}

\indent According to Eq.~\eqref{eq:eqPoint}, high values of $c$ are detrimental for vaccination, while high values of $T$ and $\beta$ boosts it. Additionally, the vaccination coverage depends exclusively on the ratio between vaccination and recovery costs, which we denote as $f= c/T$. Therefore, considering Eq.~\eqref{eq:eqPoint}, one can see that full vaccination cannot be stable for a perfect vaccine unless $f = 0$. \\
\indent More subtle is the dependence on the vaccination quality $\gamma$. In fact, lowering vaccine quality increases vaccine coverage ($dy^{*}/d\gamma > 0$) if the fraction of effectively vaccinated agents $y_{\text{eff}} > 0.5$, where $y_{\text{eff}} = (1-\gamma)y$. Noteworthy, albeit the quality of the vaccine worsens, agents will choose more often to vaccinate. 

\begin{figure}[t!]
\centering
\includegraphics[width=0.95\columnwidth]{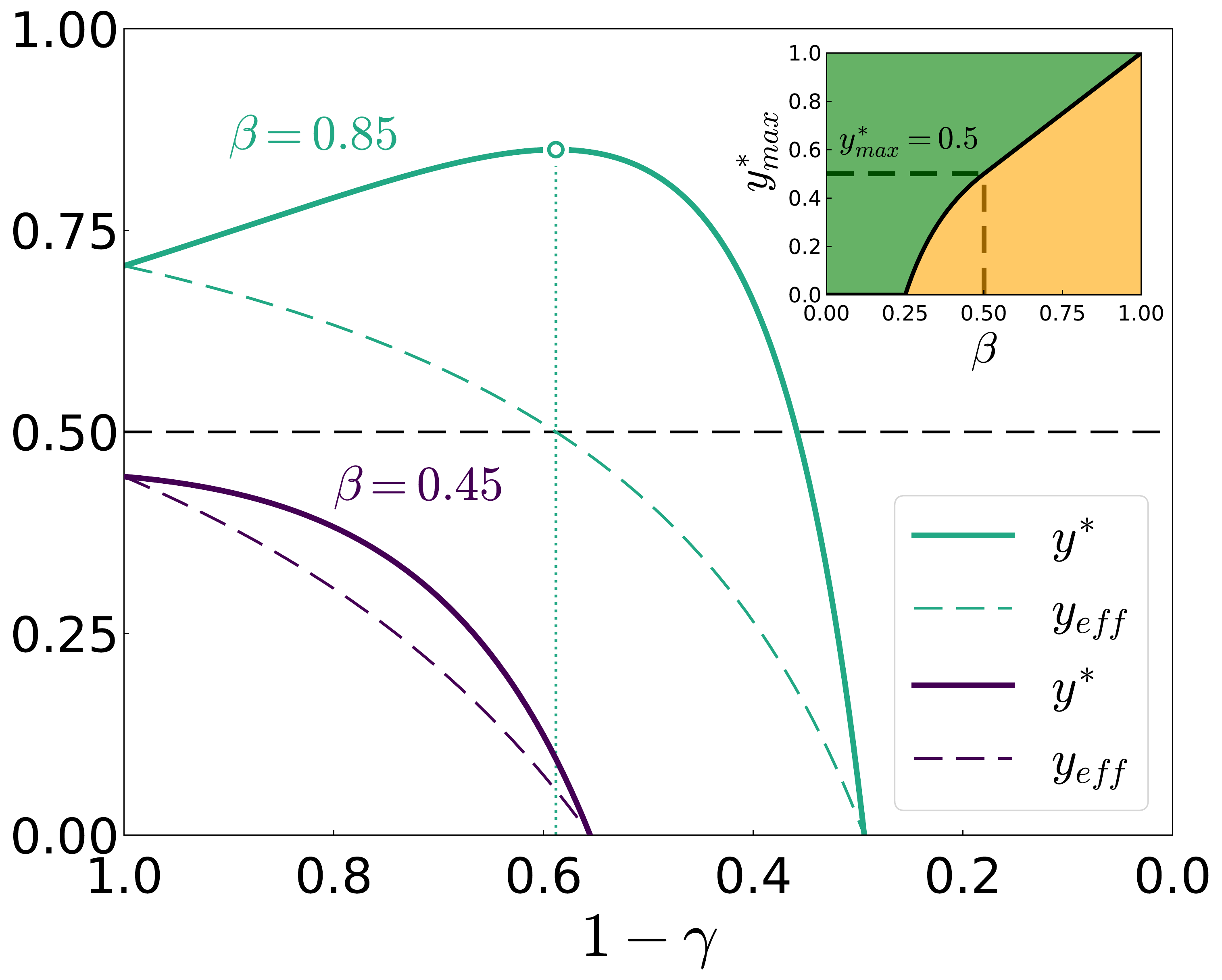}
\caption{Vaccination coverage at equilibrium, $y^*$, as a function of effectiveness, $1-\gamma$ for two different values of the infection probability $\beta$. The dashed line indicates the fraction of effectively vaccinated agents. The other parameters are $\alpha=0.4$, $c=0.1$, and $T=1.0$. The maximum coverage $y_{\max}^*$ is denoted by a point, and the dotted line delimits the tolerance range. The inset displays the value of the maximum coverage $y_{\max}^*$ given in Eq.~\eqref{eq:maxVaccination} as a function of infectivity $\beta$. The color highlights the region where vaccination takes place or not.}
\label{fig:vac_quality}
\end{figure}

At first glance, the increase in vaccine uptake as its effectiveness decrease may seem counterintuitive. The rationale behind this is as follows: as $\gamma$ increases there is a competition between the increasing risk of getting infected and the reduced protection bestowed by the vaccine itself, as shown in Fig.~\ref{fig:vac_quality}.
%
%
%
%
%
To shed light on such competition, one may look at the decrease of the vaccine's effectiveness $1-\gamma$ as a dynamical process. As effectiveness decreases, the risk for a not vaccinated agent of getting infected by a previously vaccinated one becomes $y_{\text{eff}} \times \beta \alpha \gamma$. Conversely, the risk of a vaccinated agent to get infected by a non vaccinated becomes $(1-y_{\text{eff}}) \times \beta \alpha \gamma$. Comparing these two risks, the vaccinated agent has an advantage over the non vaccinated if $y_{\text{eff}} > (1-y_{\text{eff}})$, which translates into $y_{\text{eff}} > 0.5$. Hence, the counterintuitive act of vaccinating when the efficiency of the vaccine is low, turns out to be a rational decision to mitigate the infection pressure. Moreover, the existence -- for each infectivity $\beta$ -- of a maximum fraction of vaccinated agents, $y_{\max}^*$, delimitates a region of ``tolerable effectiveness'' beyond which agents decide to non vaccinate. Also, the position of $y_{\max}^*$ versus $\beta$ splits the phase space into two distinct regions (Fig.~\ref{fig:vac_quality} inset). Interestingly, this effect has been observed previously in other coevolutionary models \cite{cardillo-pre-2013,wu-pone-2011,alvarez_zuzek-pone-2017}, but never explained hitherto. At variance with $y$, the fraction of \emph{effectively} vaccinated agents, $y_{\text{eff}}$, is always lowered by a decrease of the vaccine quality, \ie $\frac{d((1-\gamma)y^{*})}{d\gamma} < 0$. Therefore, the higher vaccination coverage is not counterbalancing the lower vaccine quality. Let us analyze this dependency more in detail.

%
\subsection{Dependence of the vaccination uptake on vaccine quality}
\label{sec:y_on_gamma}
There are three different regimes regarding the dependency of the vaccination coverage, $y_{\text{eq}}$, on $\gamma$. The first one is the absence of vaccination independently of $\gamma$, corresponding to $\beta < \tfrac{c}{\alpha T} = \tfrac{f}{\alpha}$. In the second regime $y^{*} (\gamma = 0) < 0.5$ the vaccination coverage is monotonously decreasing with $\gamma$. Therefore, maximal vaccination coverage is reached for a perfect vaccine $\gamma = 0$. The condition $y^{*} (\gamma = 0) < 0.5$ is equivalent to $\beta \leq \tfrac{2f}{\alpha}$. In the third regime $\beta > \tfrac{2f}{\alpha}$, the dependence of $y{*}$ on $\gamma$ is non monotonous, wherefore the vaccination coverage is maximal for a non perfect vaccine. The vaccine quality maximizing the vaccination coverage is then found from $dy^{*}/ d\gamma = 0$, giving:
\begin{equation}
\gamma_{c} = 1-\frac{2f}{\beta \alpha} \,.
\end{equation}

The vaccine quality maximizing the vaccination coverage can also be seen as a tolerance threshold. If $1-\gamma$ becomes worse than $1-\gamma_{c}$, agents start refusing taking the vaccine and the vaccination coverage drops rapidly, as illustrated in Fig.~\ref{fig:vac_quality}. From there, the maximal vaccination coverage in the three regimes is then given by:
\begin{equation}
\label{eq:maxVaccination}
\begin{cases}
y^{*}_{\text{max}} = 0 &\ \text{if} \ \beta \leq \frac{f}{\alpha}  \\
y^{*}_{\text{max}}  = 1-\frac{f}{\beta \alpha} &\ \text{if} \ \frac{f}{\alpha} < \beta \leq \frac{2f}{\alpha} \\
y^{*}_{\text{max}} = \frac{\beta \alpha}{4f} &\ \text{if} \ \frac{2f}{\alpha} \leq \beta  \,.
\end{cases}
\end{equation}

The maximal vaccination coverage, $y^{*}_{max}$, as a function of $\beta$ is presented, instead, in the inset of Fig.~\ref{fig:PhaseSpace}.

%
\section{Impact of vaccination on the epidemic spreading}

After discussing the outcome of the vaccination dynamics, we now turn our focus towards its impact on the subsequent epidemic outbreak. Assuming that the vaccination dynamics always ends up in the unique stable equilibrium point, $y_{\text{eq}}$; we can use such information to compute the extent of a future epidemic outbreak by using a SIR compartmental model \cite{anderson-book-1992}. The dynamics of the SIR model is then given by:
\begin{equation}
\label{eq:SIREquations}
\begin{cases}
\,\dot{S} &\!\!\!\!= -\beta IS \\
\,\dot{I} &\!\!\!\!= \beta IS -\mu I \\
\,\dot{R} &\!\!\!\!= \mu I
\end{cases}\,,
\end{equation}
where $S$, $I$, and $R$ denote the fraction of susceptible, infected, and recovered agents, respectively. The aforementioned quantities fulfill the conservation law\newline $S+I+R+(1-\gamma)y_{\text{eq}} = 1$. Note that $\mu = 1/T$ indicates the recovery probability in agreement with the vaccination game. The fraction of recovered agents after the epidemics dies out, $R_{\infty}$, is given by the following trascendental equation:
\begin{equation}
\label{eq:transcendental}
R_{\infty} = 1-(1-\gamma)y_{\text{eq}}- \Bigl[1-(1-\gamma) y_{\text{eq}}-I_0\Bigr] \, e^{-\frac{\beta}{\mu} \, R_{\infty}} \,,
\end{equation}
where $I_0$ is the initial fraction of infected agents. Eq.~\eqref{eq:transcendental} has no analytical solution, hence it must be solved numerically. Nevertheless, in an infinite size system ($N \rightarrow \infty$), the term $I_0$ can be neglected since $I_0 \ll 1$. Thus, a non negligible fraction of recovered agents, $R_{\infty}$, exists if:
\begin{equation}
\label{eq:thresholdImplicit}
\beta > \frac{\mu}{1-(1-\gamma )y_{\text{eq}}} = \beta_c \,,
\end{equation}
where $\beta_c$ denotes the epidemic threshold. As expected, if the system displays no vaccination, $y_{\text{eq}} = 0$, and the above criteria reduces to the purely epidemic SIR threshold $\beta_c = \mu$. Instead, for the full vaccination case, $y_{\text{eq}} = 1$, the criteria becomes $\beta_c = \mu/\gamma$. Finally, for the internal fixed point, $y_{\text{eq}} = y^{*}$, the existence of $R_{\infty}>0$ in Eq.\eqref{eq:transcendental} implies:
\begin{equation}
\label{eq:thresholdExplicit}
\frac{f}{\mu \alpha(1-\gamma)} > 1 \,.
\end{equation}
\indent Surprisingly, this condition is independent of $\beta$. Therefore, the increased vaccination coverage balances the increased transmission probability in the subsequent epidemic outbreak.\\
%
%
%
%
\begin{figure*}[t!]
\centering
\begin{minipage}{0.25\linewidth}
	\includegraphics[width=\linewidth]{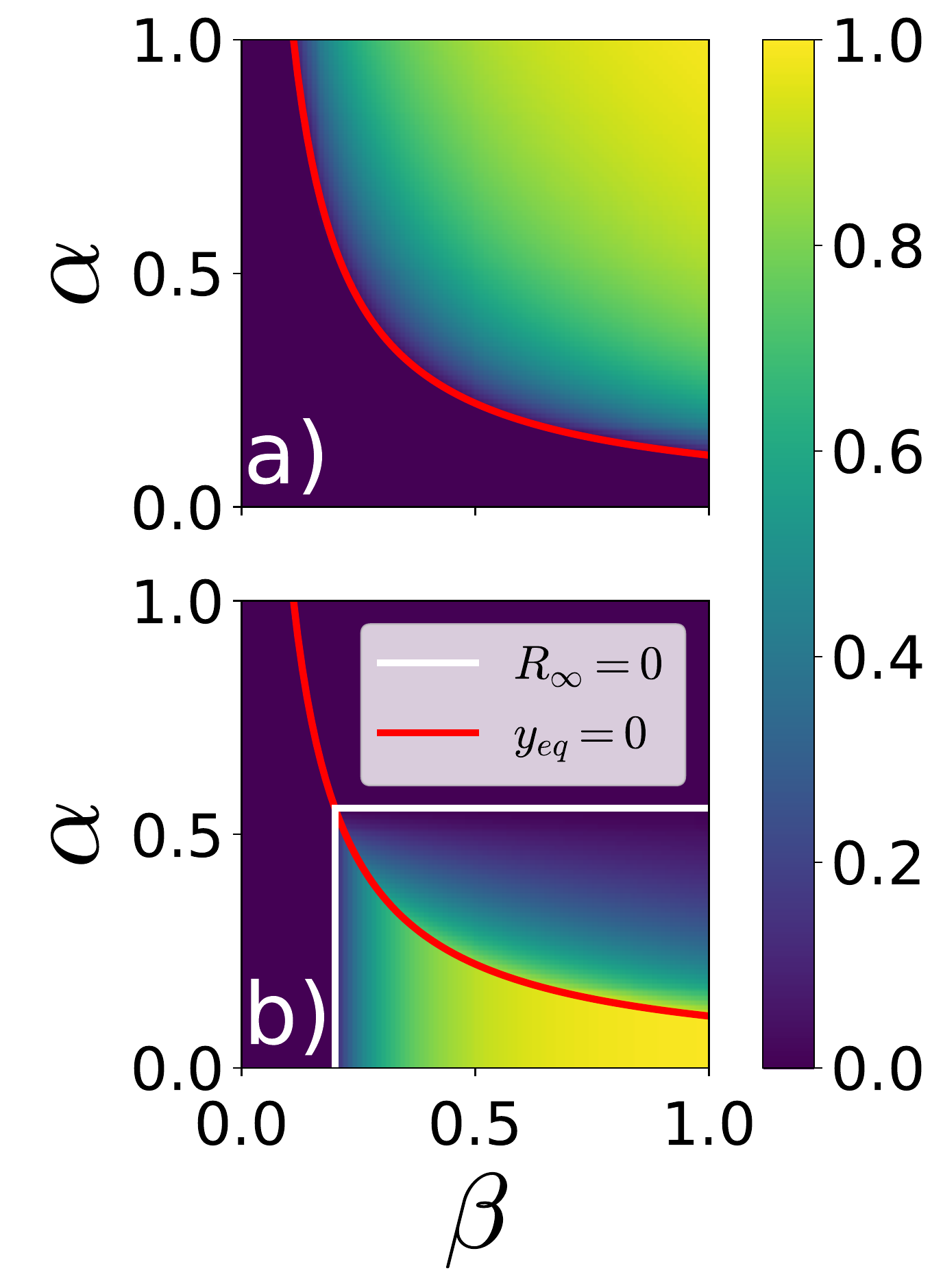}
\end{minipage}
	\begin{minipage}{0.52\linewidth}
	\includegraphics[width=\linewidth]{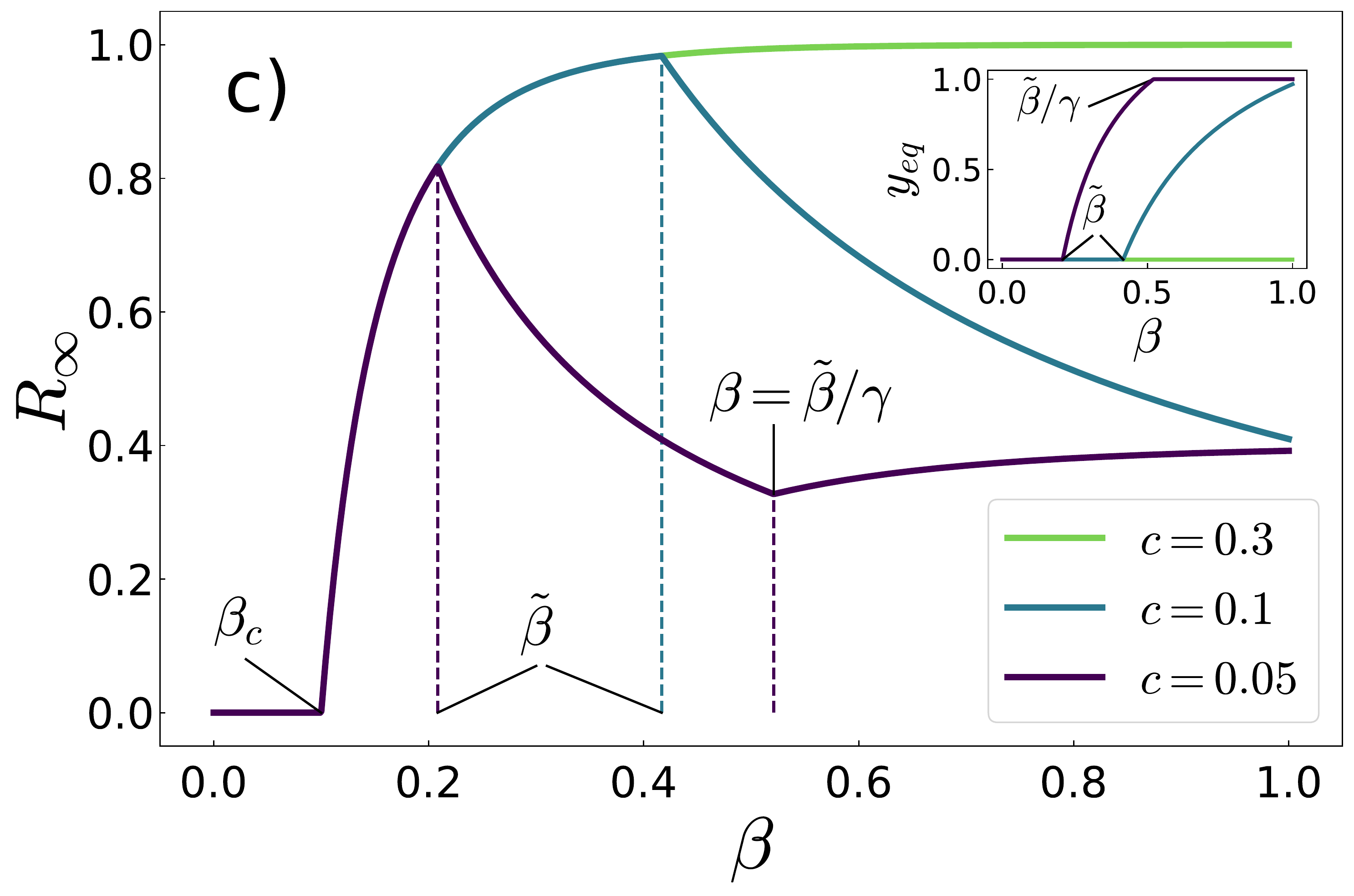}
\end{minipage}
\caption{Vaccination coverage, $y_{\text{eq}}$, {\bf (panel a) } and fraction of recovered individuals, $R_{\infty}$, {\bf(panel b)} as a function of disease incidence in the previous season, $\alpha$, and infection probability, $\beta$. The red solid line corresponds to the vaccination threshold, $\tilde{\beta}$, separating the region where no vaccine uptake occurs from the one where nonzero vaccination could be observed. In (panel b), the white solid lines correspond to the epidemic thresholds in the absence of ($\beta=0.2$), and in presence ($\alpha = 0.5$) of vaccination according to Eq.~\eqref{eq:thresholdImplicit}. The other variables are fixed to $c = 0.1$, $T = 1.0$, $\gamma = 0.1$ and $\mu = 0.2$. {\bf(panel c)} $R_{\infty}$ as a function of $\beta$ for different costs of the vaccine, $c$. The inset presents the vaccination coverage, $y_{eq}$. Additionally, the critical values of $\beta$ are presented following Eq.\eqref{eq:transcendentalCases} such as the purely epidemic threshold, $\beta_c$, the vaccination threshold, $\tilde{\beta}$, and the threshold for full vaccination, $\tilde{\beta}/\gamma$. The remaining variables are fixed to $T=1.0$, $\gamma = 0.4$, $\alpha = 0.4$ and $\mu = 0.1$.}
\label{fig:PhaseSpace}
\end{figure*}
\indent In Fig.~\ref{fig:PhaseSpace}, we display the vaccination coverage, $y_{\text{eq}}$ (panel a), and the fraction of recovered agents, $R_{\infty}$ (panel b), as a function of the previous season incidence $\alpha$, and the probability of infection $\beta$ in the case of a perfect vaccine ($\gamma = 0$). As expected, a remarkably high fraction of recovered agents is observed for a highly infective epidemic (large $\beta$) paired with a small fraction of infected agents in the previous outbreak (small $\alpha$). In (panel b), one can observe that, given a value of $\alpha$, the fraction of recovered individuals is maximal at the vaccination threshold $\tilde{\beta}$. A way of finding the maxima and minima of the fraction of recovered agents, $R_{\infty}$, is considering its derivative with respect to the tranmission probability, $\beta$, $\tfrac{d\, R_{\infty}}{d\beta}$. Note that in an infinite system, the transcendental equation Eq.~\eqref{eq:transcendental} takes three different forms depending on the vaccination coverage:  
\begin{equation}
\label{eq:transcendentalCases}
\begin{cases}
R_{\infty} = 1-e^{-\frac{\beta}{\mu}R_{\infty}} &\text{if}\quad \mu < \beta \leq \tilde{\beta}\\[0.3cm]
R_{\infty} = \frac{\tilde{\beta}}{\beta}\Bigl(1-e^{-\frac{\beta}{\mu}R_{\infty}}\Bigr) &\text{if}\quad \tilde{\beta} < \beta \leq \frac{\tilde{\beta}}{\gamma}\\[0.3cm]
R_{\infty} = \gamma \Bigl(1-e^{-\frac{\beta}{\mu}R_{\infty}}\Bigr) &\text{if}\quad \frac{\tilde{\beta}}{\gamma} < \beta \leq 1
\end{cases} \,.
\end{equation}

In order to find an explicit expression for $\tfrac{d\, R_{\infty}}{d\beta}$, we derive both sides of Eq.~\eqref{eq:transcendentalCases} and subsequently rearrange terms, which leads to:
\begin{equation}
\label{eq:derivatives}
\begin{cases}
\dfrac{dR_{\infty}}{d\beta} =\cfrac{R_{\infty}}{\mu\biggl(e^{\frac{\beta}{\mu}R_{\infty}}-\cfrac{\beta}{\mu}\biggr)} &, \ \text{if} \ \mu < \beta \leq \tilde{\beta} \\[1cm]
\dfrac{dR_{\infty}}{d\beta} =\cfrac{R_{\infty}+\cfrac{\mu}{\beta}\left(1-e^{\frac{\beta}{\mu}R_{\infty}}\right)}{\cfrac{\mu \beta}{\tilde{\beta}}\biggl(e^{\frac{\beta}{\mu}R_{\infty}} - \cfrac{\tilde{\beta}}{\mu}\biggr)} &, \ \text{if} \ \tilde{\beta} < \beta \leq \frac{\tilde{\beta}}{\gamma} \\[1cm]
\dfrac{dR_{\infty}}{d\beta} =\cfrac{R_{\infty}\gamma}{\mu\biggl(e^{\frac{\beta}{\mu}R_{\infty}}-\gamma \cfrac{\beta}{\mu}\biggr)} &, \  \text{if} \ \frac{\tilde{\beta}}{\gamma} < \beta \leq 1 \,.
\end{cases}
\end{equation} 

To infer information about the sign of $\tfrac{d\,R_{\infty}}{d\beta}$ we need to bound $R_{\infty}$. By using again Eq.~\eqref{eq:transcendentalCases} and the inequality $1\!-\!e^{-x} \!>\! \frac{x}{1+x}$, we find a lower bound for $R_{\infty}$ as:
\begin{equation}
\label{eq:lowerBounds}
\begin{cases}
R_{\infty} > 1-\frac{\mu}{\beta} &, \ \text{if} \ \mu < \beta \leq \tilde{\beta}\\
R_{\infty} > \frac{\tilde{\beta}}{\beta}-\frac{\mu}{\beta} &, \ \text{if} \ \tilde{\beta}< \beta \leq \frac{\tilde{\beta}}{\gamma} \\
R_{\infty} > \gamma-\frac{\mu}{\beta} &, \  \text{if} \ \frac{\tilde{\beta}}{\gamma} < \beta \leq 1 \,.
\end{cases}
\end{equation}

These lower bounds can then be combined with the inequality $e^{\frac{\lambda}{\mu}R_{\infty}} \geq 1+\frac{\lambda}{\mu}R_{\infty}$. The strict inequality holds for all $R_{\infty} > 0$. This enables us to develop the expressions for the derivative in Eq.~\eqref{eq:derivatives}, leading to:
\begin{equation}
\label{eq:sign}
\begin{cases}
\frac{dR_{\infty}}{d\beta} > 0 &, \ \text{if} \ \mu < \beta \le \tilde{\beta} \\
\frac{dR_{\infty}}{d\beta} < 0 &, \ \text{if} \ \tilde{\beta} < \beta \le \frac{\tilde{\beta}}{\gamma} \\
\frac{dR_{\infty}}{d\beta}  > 0 &, \  \text{if} \ \frac{\tilde{\beta}}{\gamma} < \beta \le 1 \,.
\end{cases}
\end{equation}

The cases $\beta < \tilde{\beta} $ and $\frac{\tilde{\beta}}{\gamma} < \beta$ are as expected. In these regimes the vaccination coverage does not vary with $\beta$ ($y_{\text{eq}}=0$ and $y_{\text{eq}} = 1$, respectively), wherefore $R_{\infty}$ increases monotonously with $\beta$. In the intermediate regime $\tilde{\beta} < \beta < \tilde{\beta}/\gamma$ though, where $y_{\text{eq}}= y^{*}$, $R_{\infty}$ monotonously decreases with $\beta$ ($\frac{dR_{\infty}}{d\beta} <0$). Consequently, vaccination emerges in a way such that it outweighs the increased transmission probability and hinders the epidemic spreading, as illustrated in Fig.\ref{fig:PhaseSpace}c. From these considerations, we can conclude that the fraction of recovered agents is locally maximal at $\beta = \tilde{\beta}$, and locally minimal at $\beta = \tilde{\beta}/\gamma$. The numerical exploration of the parameter space has confirmed hitherto that the former is also a global maximum. Moreover, the highest fraction of recovered, $R_{\infty}^{\text{max}}$, for a fully vaccinated population ($y_{\text{eq}} = 1$) would correspond to meaningless values of the parameters. Depending on the parameters, the system will not be in all of the three regimes presented in Eq.~\eqref{eq:sign} as the transmission probability, $\beta$, is varied. If the parameters are such that $\tilde{\beta}/\gamma > 1$, the system will never fall in the third regime. Consequently, $R_{\infty}$ monotonously decreases once vaccination emerges. Similarly, one might have $\tilde{\beta} > 1$ for which the system shows no vaccination and $R_{\infty}$ monotonously increases with $\beta$. The different cases are illustrated in Fig.\ref{fig:PhaseSpace}c.

%
\subsection{Numerical simulations}

\begin{figure}[ht!]
\centering
\includegraphics[width=0.99\columnwidth]{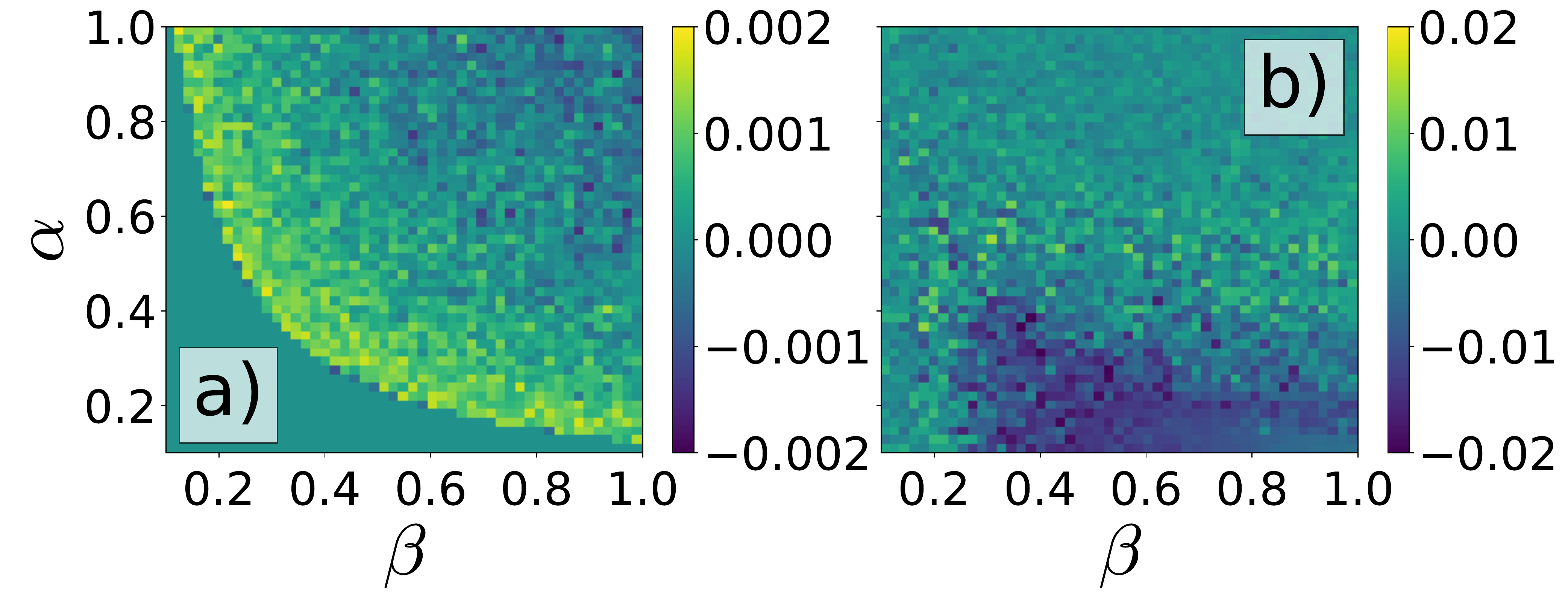}
\caption{Differences between the quantities computed using Monte Carlo simulations and their analytical counterparts as a function of parameters $\alpha$ and $\beta$. Panel $a$ presents the case of $y_{\text{eq}}$, while panel $b$ the case of $R_\infty$, instead. The other parameters are set to $N = 1000$, $\gamma=0.1$, $\mu = 0.2$, and $T=1.0$, respectively.}
\label{fig:compare}
\end{figure}

To validate that the previous equations are actually describing the behavior of the considered system, we have compared the analytical results discussed above with those obtained via Monte Carlo simulations. The simulations have been made for a system of $N = 1000$ agents updating their strategy accordingly to Eq.~\eqref{eq:updateRule}, and averaged over $N_{real} = 50$ realizations. The difference between theory and the simulation for the vaccine coverage, $y$, and fraction of infected agents, $R_{\infty}$, is plotted in panels $a$ and $b$ of Fig.~\ref{fig:compare}. The maximal differences for $y$ and $R_{\infty}$ are around $0.2 \%$ and $2.5 \%$, respectively. The average relative errors, instead, are $0.2 \%$ for $y$ and $2.0  \%$ for $R_{\infty}$. 
%
%
%
The agreement between analytical and numerical simulations, $\sigma$, is thus $\sim 2\%$.\newline
\indent We want to remark also that preliminary analysis made in the case of discrete interactions encoded as networks showed similar results, although we lack an analytical solution. Notwithstanding, the well-mixed population proves to be a good approximation for the vaccination dynamics on networks. This is very likely due to the vaccination cost term in the total payoff, which does not scale with the degree of nodes. Additionally, the nontrivial increase in vaccine uptake -- as effectiveness decreases -- is also observed in simulations on networked populations \cite{cardillo-pre-2013}. \newline

%
\section{Conclusions}

Summing up, we have presented a mean-field model to predict vaccine uptake for influenza-like diseases using the disease incidence during previous outbreak season as a proxy for the ``perception of infection risk.'' The model predicts the existence of a tolerance range of vaccine effectiveness, where a nontrivial increase in vaccination coverage takes place as the vaccine inefficiency increases. Albeit appearing irrational and counterintuitive at first sight, such behavior is -- instead -- due to the interplay between the vaccination game and the disease spreading processes. The model predicts also that highly infective -- but under control -- epidemics might prove dangerous for future infections, since they alter the ``risk perception'' of the agents, and induce them to non vaccination. Finally, as mentioned in the introduction, the time scale between the two dynamics (decision and spreading) has always been considered fixed. The sole exception, up to our knowledge, are childhood diseases with long term immunity \cite{bauch-proc_roy_soc_b-2005,bauch-plos_comp_bio-2012}. The framework presented here could be used to fill the gap among different model formulations and unify them.

\begin{acknowledgements}
PDLR acknowledges the financial support of Swiss National Science Foundation under Grant No. CRSII2\_147609. AC acknowledges the financial support of MINECO through Grant RYC-2012-01043. JGG acknowledges financial support from MINECO (projects FIS2015-71582-C2 and 377 FIS2014-55867-P) and from the Departamento de Industria e Innovaci\'on del Gobierno de Arag\'on y Fondo Social Europeo (FENOL group E-19). AA acknowledges financial support from Spanish MINECO (grant FIS2015-71582-C2-1), Generalitat de Catalunya ICREA Academia, and the James S. McDonnell Foundation.
\end{acknowledgements}


\end{document}